\documentclass[twocolumn]{IEEEtran} 

\usepackage{graphicx}
\usepackage{xspace}

\def\BibTeX{{\rm B\kern-.05em{\sc i\kern-.025em b}\kern-.08em
    T\kern-.1667em\lower.7ex\hbox{E}\kern-.125emX}}

\newcommand{\gsim}{{\scriptscriptstyle\stackrel{>}{\sim}}}
\newcommand{\lsim}{{\scriptscriptstyle\stackrel{<}{\sim}}}

\begin{document}

\title{Voltage-flux-characteristics of asymmetric dc SQUIDs\\} 

\author{Jochen M\"{u}ller, Stefan Weiss, Rudolf Gross, Reinhold Kleiner, Dieter Koelle
\thanks{Manuscript received September 17, 2000.}
\thanks{J. M\"{u}ller was with
II. Physikalisches Institut, Lehrstuhl f\"{u}r Angewandte
Physik, Universit\"{a}t zu K\"{o}ln, Z\"{u}lpicherstr. 77, D-50937 K\"{o}ln,
Germany. He is now with RWTH Aachen, Lehrstuhl f\"{u}r
Prozessleittechnik, Turmstrasse 46, D-52064 Aachen,
Germany. E-mail: JOCHEN@plt.rwth-aachen.de .}
\thanks{S. Weiss was with
II. Physikalisches Institut, Lehrstuhl f\"{u}r Angewandte
Physik, Universit\"{a}t zu K\"{o}ln, Z\"{u}lpicherstr. 77, D-50937 K\"{o}ln,
Germany. He is now with ESAS, Stuttgart, Germany. E-mail: weiss.stephan@web.de .}
\thanks{R. Gross was with
II. Physikalisches Institut, Lehrstuhl f\"{u}r Angewandte Physik,
Universit\"{a}t zu K\"{o}ln, Z\"{u}lpicherstr. 77, D-50937 K\"{o}ln, Germany. He
is now with Walther-Mei{\ss}ner-Institut, Bayerische Akademie der
Wissenschaften, Walther-Mei{\ss}ner Str. 8, D-85748 Garching,
Germany. E-mail: rudolf.gross@wmi.badw.de .}
\thanks{R. Kleiner is with
Physikalisches Institut -- Experimentalphysik II, Universit\"{a}t T\"{u}bingen,
Morgenstelle 14, D-72076 T\"{u}bingen, Germany. E-mail:
kleiner@uni-tuebingen.de .}
\thanks{D. Koelle is with
II. Physikalisches Institut, Lehrstuhl f\"{u}r Angewandte
Physik, Universit\"{a}t zu K\"{o}ln, Z\"{u}lpicherstr. 77, D-50937 K\"{o}ln,
Germany. E-mail: koelle@ph2.uni-koeln.de .}
}

\markboth{IEEE Transactions On Applied Superconductivity, Vol. 10 (2001)
\hspace{5mm} ASC 2000, Virginia Beach (USA) --  4ED02}
{M\"{u}ller {\it et al.}: Voltage-flux-characteristics of asymmetric dc SQUIDs} 


\maketitle

\begin{abstract}
We present a detailed analysis of voltage-flux
$V(\Phi)$-characteristics for asymmetric dc SQUIDs with
various kinds of asymmetries. For finite asymmetry
$\alpha_I$ in the critical currents of the two
Josephson junctions, the minima in the
$V(\Phi)$-characteristics for bias currents of opposite
polarity are shifted along the flux axis by $\Delta\Phi
= \alpha_I \beta_L$ relative to each other; $\beta_L$
is the screening parameter. This simple relation allows
the determination of $\alpha_I$ in our experiments on
$\rm YBa_2Cu_3O_{7-\delta}$ dc SQUIDs and comparison
with theory.  Extensive numerical simulations within a
wide range of $\beta_L$ and noise parameter $\Gamma$
reveal a systematic dependence of the transfer function
$V_\Phi$ on $\alpha_I$ and $\alpha_R$ (junction
resistance asymmetry) . As for the symmetric dc SQUID,
$V_\Phi$ factorizes into $g(\Gamma\beta_L)\cdot
f(\alpha_I,\beta_L)$, where now $f$ also depends on
$\alpha_I$. For $\beta_L\lsim 5$ we find mostly a decrease of $V_\Phi$ with
increasing $\alpha_I$, which however can only partially
account for the frequently observed discrepancy in
$V_\Phi$ between theory and experiment for high-$T_c$
dc SQUIDs.
\end{abstract}


\section{Introduction}

\PARstart{T}{he} observation of a significant
discrepancy between numerical simulations and
experimental results obtained for direct current (dc)
superconducting quantum interference devices (SQUIDs)
based on high-transition-temperature superconductors
(HTS) is one of the most important unsolved problems for
HTS dc SQUIDs which seriously hinders their optimization
for applications \cite{koelle99}. HTS dc SQUIDs show
frequently asymmetric behavior which may be attributed
to the large spread in the critical current $I_0$ and
normal resistance $R$ of HTS Josephson junctions. This
may lead to asymmetric critical current $I_c$ or voltage
$V$ vs. external flux $\Phi$ characteristics of the dc
SQUID \cite{kleiner96,tesche77} and can affect the transfer function
$V_\Phi\equiv |dV/d\Phi|_{max}$ which is defined as the
maximum slope of the $V(\Phi)$-curves.
However, such an asymmetry has been usually neglected in
numerical simulations of $V_\Phi$ for HTS dc SQUIDs.

In this paper we present a detailed study of the impact
of asymmetry on the $V(\Phi)$-characteristics and in
particular on the transfer function of dc SQUIDs.
We first introduce the main parameters which define the
asymmetric dc SQUID (Sec.\ref{asymmetric dc SQUID}).
Then we show that an asymmetry in $I_0$ of the two
Josephson junctions leads to a shift of the
$I_c(\Phi)$- and in the $V(\Phi)$-characteristics,
which can be used to determine the critical current
asymmetry experimentally, as demonstrated on dc SQUIDs
with $\rm YBa_2Cu_3O_{7-\delta}$ (YBCO) bicrystal grain
boundary Josephson junctions \cite{barthel99}
(Sec.\ref{critical current vs. flux and voltage vs.
flux characteristics}). We present numerical simulation
results for $V_\Phi$ which we obtained within a wide
range of parameters including the limit of large
thermal fluctuations which are important for HTS dc
SQUIDs (Sec.\ref{transfer function: numerical
simulations}), and we compare those results with
experimental data on HTS dc SQUIDs (Sec.\ref{transfer
function: numerical simulation vs. experiment}).

\section{asymmetric dc SQUID}
\label{asymmetric dc SQUID}

\begin{figure}[2b]
\center{\includegraphics [width=0.5\columnwidth,clip]
{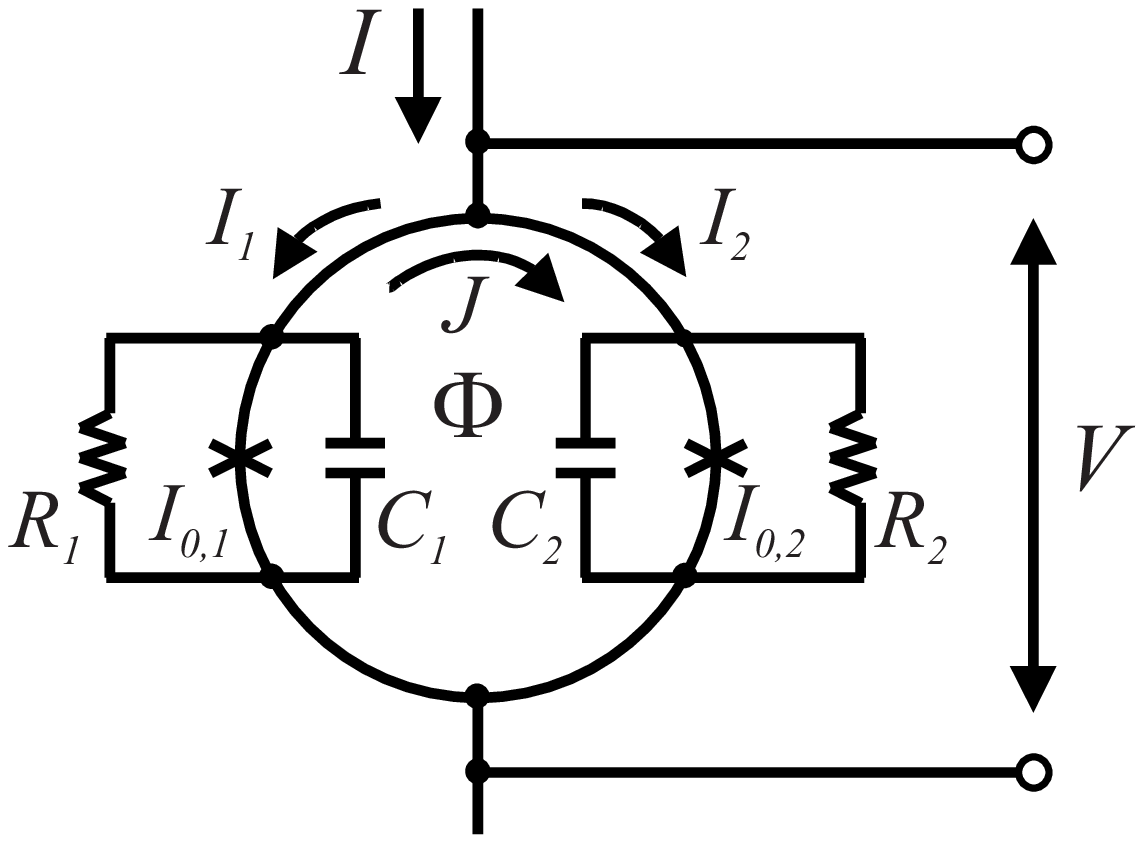}}
\caption{The asymmetric dc SQUID}
\label{f-esasymsq}
\end{figure}

The asymmetric dc SQUID shown in Fig.\ref{f-esasymsq}
consists of a superconducting loop of inductance $L$
intersected by two Josephson junctions with average
values of critical current $I_0$, resistance $R$ and
capacitance $C$.
The asymmetry in the junction
parameters is described via the asymmetry parameters $\alpha_I$,
$\alpha_R$ and $\alpha_C$, wich are defined according to
\begin{equation}
\begin{array}{lll}
I_{0,1}=I_0(1-\alpha_I), &R_{1}=R/(1-\alpha_R), &C_{1}=C(1-\alpha_C),\\[2mm]
I_{0,2}=I_0(1+\alpha_I), &R_{2}=R/(1+\alpha_R), &C_{2}=C(1+\alpha_C),
\end{array}
\end{equation}
where the subscripts 1,2 denote the parameters of the
left and right junction, respectively.

Throughout this paper we consider two different origins
of the junction asymmetry, noting, that in real devices
we may have a combination of both: (i) geometric
asymmetry and (ii) intrinsic asymmetry.

In the case of
geometric asymmetry we assume identical values of
critical current density $j_0\equiv I_0/A$, resistance
times area $\rho\equiv R\times A$, and specific
capacitance $C'\equiv C/A$ for both junctions. Here,
$A=w\times l$ is the junction area with width $w$ and
length $l$. We then introduce an asymmetry via different
values for $w$, assuming constant $l$. In the case of
bicrystal grain boundary junctions $l$ equals the film
thickness $d$, which can be assumed to be constant in
practical devices. The geometric asymmetry is then
described by the asymmetry parameter $\alpha_g$
according to
$w_{1}=(1-\alpha_g)w$ and $w_{2}=(1+\alpha_g)w$,
with $w\equiv (w_1+w_2)/2$, and we find the simple
relation for the asymmetry parameters
$\alpha_g=\alpha_I=\alpha_R=\alpha_C$.

In the case of intrinsic asymmetry we assume
$\alpha_g=0$ and different values of $j_0$ and $\rho$
for the two junctions which may reflect the natural
spread in junction parameters. For simplicity we neglect
the spread in $C'$ which is expected to be much smaller
than the spread in $j_0$ and $\rho$.
The intrinsic asymmetry is then described by the
asymmetry parameters $\alpha_j$ and $\alpha_\rho$
according to
\begin{equation}
\begin{array}{ll}
j_{0,1}=j_0(1-\alpha_j), & \rho_{1}=\rho/(1-\alpha_\rho),\\[2mm]
j_{0,2}=j_0(1+\alpha_j), & \rho_{2}=\rho/(1+\alpha_\rho),
\end{array}
\end{equation}
and we get
$\alpha_I=\alpha_j$,
$\alpha_R=\alpha_\rho$, and
$\alpha_C\approx 0$.

If we assume the scaling relation $I_0R\equiv
j_0\rho\propto j_0^{1/2}$ as derived from the
intrinsically shunted junction model \cite{gross97}
we derive the relations
\begin{equation}
\alpha_\rho=[1-(1-\alpha_j^2)^{1/2}]/\alpha_j\quad
{\rm or} \quad
\alpha_j=2\alpha_\rho/(1+\alpha_\rho^2).
\label{eq-isj-correlation}
\end{equation}

Finally, if we consider a combination of geometric and intrinsic
asymmetry we can derive from the definitions of the
asymmetry parameters given above the following relations
between the asymmetry parameters
\begin{equation}
\begin{array}{lll}
\alpha_j=\frac{\alpha_g-\alpha_I}{\alpha_I\alpha_g-1},
& {\rm and}
& \alpha_\rho=\frac{\alpha_g-\alpha_R}{\alpha_R\alpha_g-1}\quad.
\end{array}
\label{eq-alpha_Ij_Rrho-correlation}
\end{equation}

\section{critical current vs. flux and voltage vs.
flux characteristics}
\label{critical current vs. flux and voltage vs.
flux characteristics}

We derive now a very simple expression for $\alpha_I$,
which can be used to determine its value experimentally
without cutting the SQUID loop . Let us first consider
the $I_c(\Phi)$-characteristics of the dc SQUID [c.f.
Fig.\ref{f-esasymsq}] for $\Gamma=0$: the maximum critical
$I_c^{max}=I_{0,1}+I_{0,2}=2I_0$ is maintained when
$I_1=I_{0,1}$ and $I_2=I_{0,2}$. In this case the
circulating current $J$ is given as
$J(I_c^{max})=(I_{0,2}-I_{0,1})/2=\alpha_II_0$, which is
flowing in the SQUID loop if we apply an external flux
$\Phi^+=\alpha_II_0L$. Hence, the maxima of the
$I_c(\Phi)$-characteristics are shifted by $\Phi^+$
along the $\Phi$-axis, as compared to the symmetric
SQUID with $\alpha_I=0$, where $I_c$ is maximum at
$\Phi=0$. The same argument leads to a negative shift
$\Phi^-=-\alpha_II_0L$ if the current direction is
reversed. Hence, the maxima of the
$I_c(\Phi)$-characteristics for opposite polarity of
the current are shifted by
$\Delta\Phi\equiv\Phi^+-\Phi^-=2\alpha_II_0L$. Using
the screening parameter $\beta_L\equiv2I_0L/\Phi_0$ we
arrive at the very simple relation

\begin{equation}
\label{eq-DeltaPhi}
\Delta\Phi/\Phi_0=\alpha_I\beta_L,
\end{equation}
which is also valid for finite values of $\Gamma$ and for
the shift of the minima of the
$V(\Phi)$-characteristics measured at constant bias
current $I\approx 2I_0$ and $-I\approx -2I_0$.

\begin{figure}[t]
\center{\includegraphics [width=0.95\columnwidth,clip]
{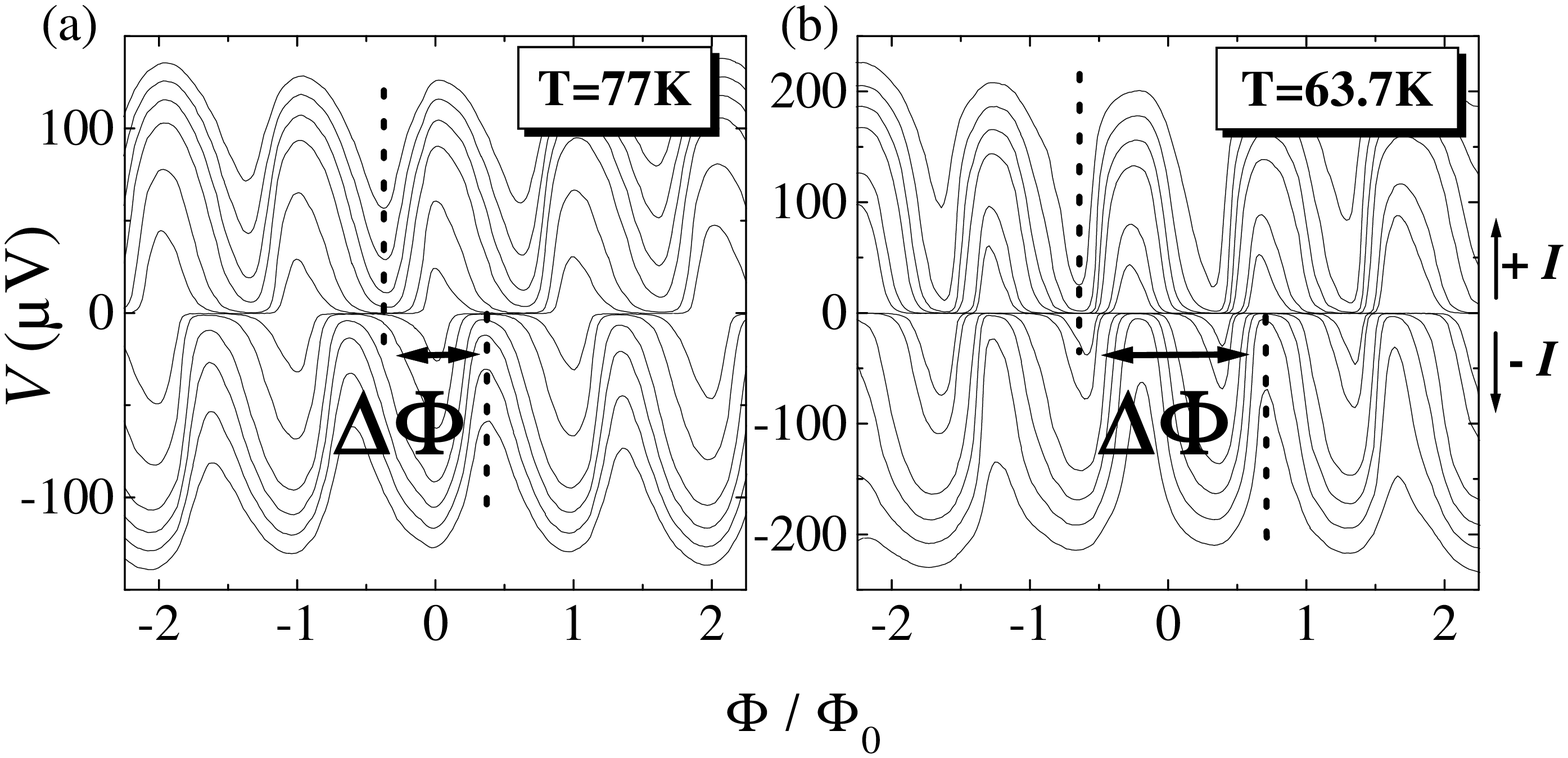}}
\caption{Measured $V(\Phi)$-curves for
different values of bias current $I$ (positive and
negative) at (a) $T=77$K and (b) $T=63.7$K.}
\label{f-V_Phi-curves}
\end{figure}
The clear-cut experimental determination of $\Delta\Phi$
requires the measurement of $V(\Phi)$-curves at various
values of $\beta_L$, which can be varied by temperature.
An example of such a measurement on one device
at $T=77$K and 63.7K is shown in
Fig.\ref{f-V_Phi-curves}.
\begin{figure}[2b]
\center{\includegraphics [width=0.75\columnwidth,clip]
{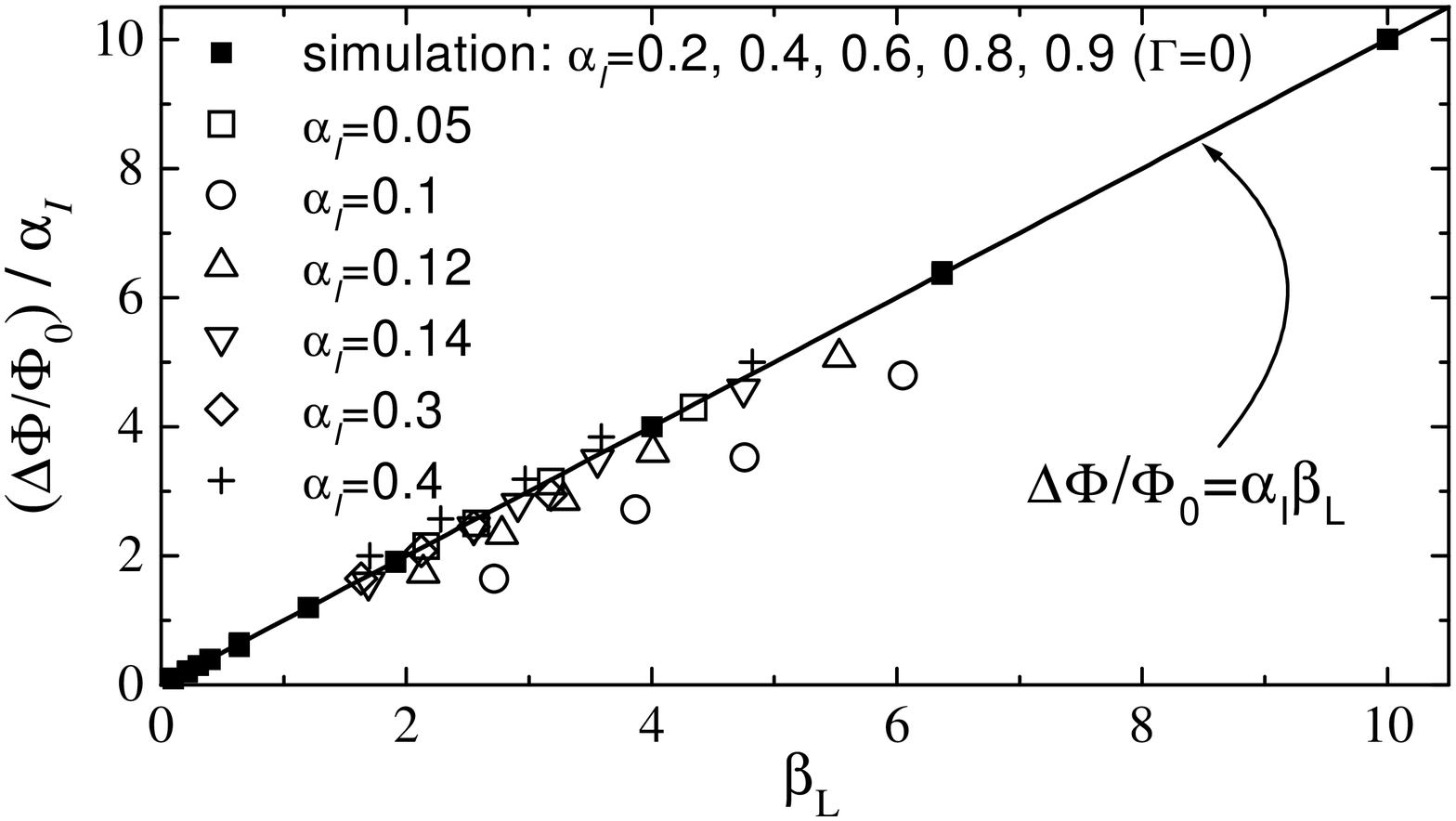}}
\caption{Measured normalized flux shift vs. $\beta_L$
for various dc SQUIDs.}
\label{f-DPhi_beta}
\end{figure}
Plotting the measured flux shift, normalized by
$\alpha_I\Phi_0$ vs. $\beta_L$ should give according to
Eq.(\ref{eq-DeltaPhi}) a linear dependence with slope 1,
with the reasonable assumption that $\alpha_I$ does not
depend on $T$.
Fig.\ref{f-DPhi_beta} shows the results of such
measurements obtained for 6 different dc SQUIDs, where
$\alpha_I$ was obtained as a fitting parameter to
give the expected slope of 1.
For comparison, the
results from simulated $V(\Phi)$-curves for various
values of $\alpha_I$ are also shown, which are in
excellent agreement with Eq.(\ref{eq-DeltaPhi}).

Except for one device, these SQUIDs have intentionally
been fabricated with a geometric asymmetry
($\alpha_g\neq 0$). From the known value of $\alpha_g$
and the measured value of $\alpha_I$ the asymmetry
parameter $\alpha_j$ can be calculated using
Eq.(\ref{eq-alpha_Ij_Rrho-correlation}). The results are
listed in Table~I.
As a main result, we see that 3 SQUIDs show only a small
asymmetry in $j_0$ with $|\alpha_j|\le0.1$. However, for
the three other devices the asymmetry in $j_0$ is
significant with values of $|\alpha_j|$ up to 0.4, which
demonstrates that the difference in critical current
density for the two junctions can be quite large.
\begin{table}[b]
\caption{Asymmetry parameters for various YBCO dc
SQUIDs.}
\begin{center}
\begin{tabular}{|c||c|c|c|c|c|c|}\hline
\# & 1 & 2 & 3 & 4 & 5 & 6 \\ \hline\hline $\alpha_g$ &
0   & 0.33  & 0.33   & 0.33 & 0.5  & 0.5   \\ \hline
$\alpha_I$ & 0.1 & 0.05  & 0.3    & 0.4  & 0.12 & 0.14
\\ \hline $\alpha_j$ & 0.1 & -0.29 & -0.04  & 0.08 &
-0.4 & -0.39 \\ \hline
\end{tabular}
\end{center}
\end{table}

\section{transfer function: numerical simulations}
\label{transfer function: numerical simulations}

As already evident from Fig.\ref{f-V_Phi-curves} the
asymmetry can induce distortions of the
$V(\Phi)$-curves, which leads to different values of
$V_\Phi^+$ and $V_\Phi^-$ for the maximum positive and
negative slope of the $V(\Phi)$-curves, respectively. To
understand the impact of the asymmetry on the transfer
function we performed numerical simulations to solve
the equations for the phase differences $\delta_1(t)$
and $\delta_2(t)$ of the two junctions \cite{kleiner96,tesche77}
over a wide range of values for $\beta_L$ and the noise
parameter $\Gamma\equiv2\pi k_BT/I_0\Phi_0$ for both,
the geometric and intrinsic asymmetry. In the latter
case we assume the correlation between $\alpha_j$ and
$\alpha_\rho$ as given in Eq.(\ref{eq-isj-correlation}).

\begin{figure}[1b]
\center{\includegraphics [width=0.95\columnwidth,clip]
{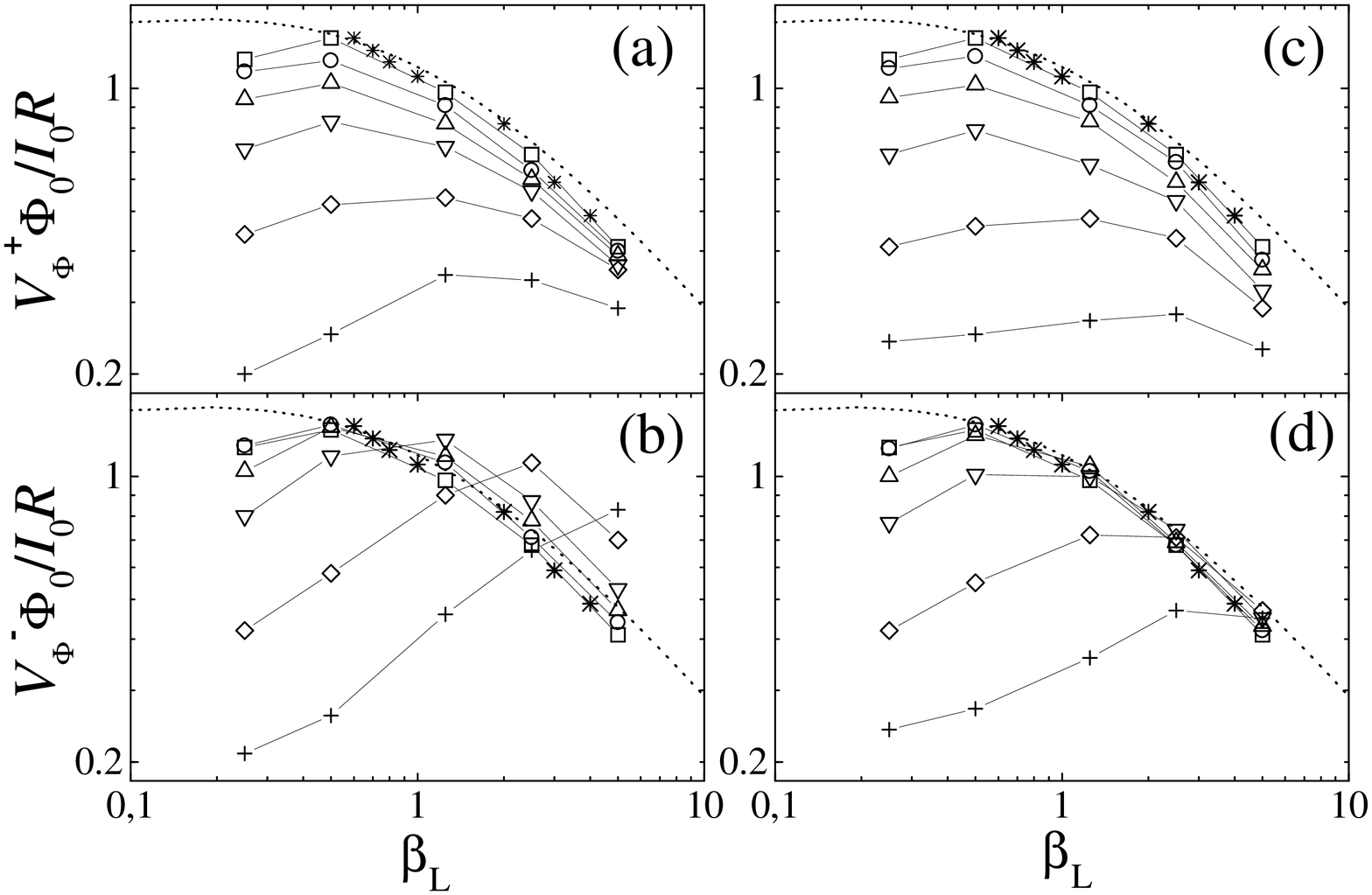}}
\caption{Calculated normalized transfer function
$V_\Phi^+$ and $V_\Phi^-$
vs. $\beta_L$ for $\Gamma\beta_L=0.2$ with
geometric asymmetry (a), (b) and intrinsic asymmetry
(c), (d); $\alpha_I$ = 0 ($\Box$), 0.2
($\scriptscriptstyle{\bigcirc}$), 0.4 ($\bigtriangleup$), 0.6
($\bigtriangledown$), 0.8 ($\Diamond$), 0.9 ($+$);
solid lines are guide to the eye.
For comparison, simulation data ($\ast$) together with
the according fit-function (dotted line) from [1] 
for symmetric dc SQUIDs are also shown.}
\label{f-Vphi_beta-sim_Gb02}
\end{figure}

Figure \ref{f-Vphi_beta-sim_Gb02} shows $V_\Phi^+$ and
$V_\Phi^-$ vs. $\beta_L$ obtained for fixed
$\Gamma\beta_L\equiv L/L_{th}=0.2$ for geometric and
intrinsic asymmetry.
\begin{figure}[t2bh]
\center{\includegraphics [width=0.92\columnwidth,clip]
{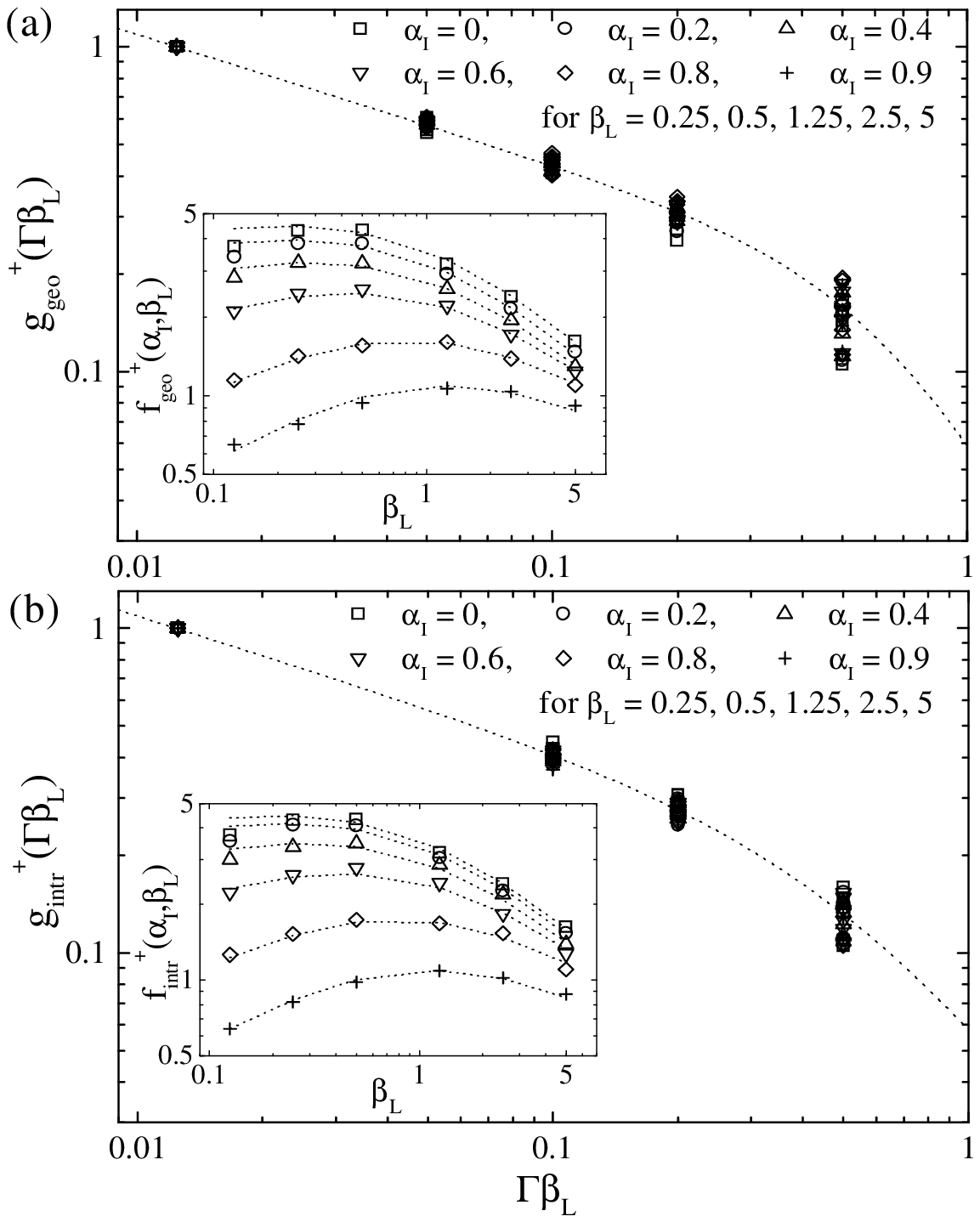}}
\caption{Calculated normalized transfer function $V_\Phi^+$ for
(a) geometric and (b) intrinsic asymmetry.
The plot shows the function $g(\Gamma\beta_L)$
which does not depend on $\alpha_I$.
The inset shows the normalization function
$f(\alpha_I,\beta_L)$.
Dashed lines are calculated from Eq.(\ref{eq-gf-plus}).}
\label{f-Vphi-norm}
\end{figure}
For $\alpha_I=0$ we closely reproduced the results
obtained in \cite{koelle99} for symmetric dc SQUIDs. In
most cases we find for $\beta_L\lsim 5$ for the
asymmetric SQUID a reduction of $V_\Phi$ as compared to
the symmetric SQUID , which increases with decreasing
$\beta_L$ and increasing $\alpha_I$. However, in the case of
geometric asymmetry, we find for intermediate values of
$1\gsim\beta_L\gsim 5$ an increase in
$V_\Phi^-$ for $\alpha_I$=$\alpha_R \gsim 0.5$, which is
concomitant with a strong distortion of the
$V(\Phi)$-characteristics. If we assume $I_0R\propto
j_0^{1/2}$, we find that for geometric asymmetry $\alpha_R$ is
always larger than for intrinsic asymmetry (for given
$\alpha_I$). This implies that the distortion in
$V(\Phi)$ is dominated by the asymmetry in the junction
resistances which becomes important for $\beta_L\gsim
1$. At $\beta_L\lsim 0.2$ the reduction of $V_\Phi^+$
and $V_\Phi^-$ is similar for both types of asymmetry,
indicating that for small $\beta_L$ the asymmetry in
the critical currents gives the main contribution to
$V_\Phi$.

Results similar to those shown in
Fig.\ref{f-Vphi_beta-sim_Gb02} have been obtained over a
wide range $1/80\le\Gamma\beta_L\le0.5$ which
corresponds to 4pH$\le L\le$160pH for the SQUID
inductance at $T$=77K where $L_{th}=321$pH. We note that
for the symmetric dc SQUID it was shown in
\cite{koelle99} that the normalized transfer function
$v_\phi\equiv V_\Phi\Phi_0/I_0R$ factorizes in
$v_\phi=g(\Gamma\beta_L)\cdot f(\beta_L)$ where
$f(\beta_L)$ is given as
$v_\phi(\beta_L;\Gamma\beta_L=1/80)$.
As a main result
of our simulations for the asymmetric dc SQUID we find a
similar factorization, with $f(\alpha_I,\beta_L)$ being
now also dependent on $\alpha_I$, while
$g(\Gamma\beta_L)$ shows no dependence on $\alpha_I$.
For $V_\Phi^+$ this is shown in Fig.\ref{f-Vphi-norm}(a)
for geometric and in Fig.\ref{f-Vphi-norm}(b) for
intrinsic asymmetry.
The simulation data shown in Fig.\ref{f-Vphi-norm} can be
approximated as

\begin{eqnarray}
g^+_{geo}(\Gamma\beta_L)&
= &
[(80 \Gamma\beta_L)^{0.4} + 0.35(4\Gamma\beta_L)^{2.5}]^{-1}
\nonumber\\[2mm]
f^+_{geo}(\alpha_I,\beta_L)&
=&
\!\!\!\frac{(7.18\cdot 0.31^{\alpha_I^{8.42}}\cdot
0.45^{\alpha_I^{1.22}})\beta_L^{0.18+0.49\alpha_I^{2.68}}}{(1 + \beta_L)}
\nonumber\\[2mm]
g^+_{intr}(\Gamma\beta_L)&
=&
[(80 \Gamma\beta_L)^{0.4} + 0.91(4\Gamma\beta_L)^{1.83}]^{-1}
\nonumber\\[2mm]
f^+_{intr}(\alpha_I,\beta_L)&
=&
\!\!\!\frac{(7.18\cdot 0.23^{\alpha_I^{11.53}}\cdot
0.41^{\alpha_I^{1.7}})\beta_L^{0.18+0.46\alpha_I^{2.55}}}{(1 + \beta_L)}
\nonumber
,\\[2mm]
\label{eq-gf-plus}
\end{eqnarray}
which is plotted in Fig.\ref{f-Vphi-norm}
as dotted lines. These equations enable one to
calculate $V_\Phi^+$ immediately for any value of
$\beta_L$, $\Gamma\beta_L$ and $\alpha_I$,
however {\it only} within the
range of parameters displayed in Fig.\ref{f-Vphi-norm}.
Similar behavior was found for $V_\Phi^-$, again with
the factorization $v_\phi^-=g^-(\Gamma\beta_L)\cdot
f^-(\alpha_I,\beta_L)$ with $g^-\approx g^+$. For $f^-$
we did not derive expressions as for $f^+$ given in
Eqs(\ref{eq-gf-plus}). As discussed above for $f^-$ the
resistance asymmetry plays an important role. Its
impact on $V_\Phi$ needs to be studied in more detail.

\section{transfer function: numerical simulation vs.
experiment}
\label{transfer function: numerical simulation vs.
experiment}

\begin{figure}[b]
\center{\includegraphics [width=\columnwidth,clip]
{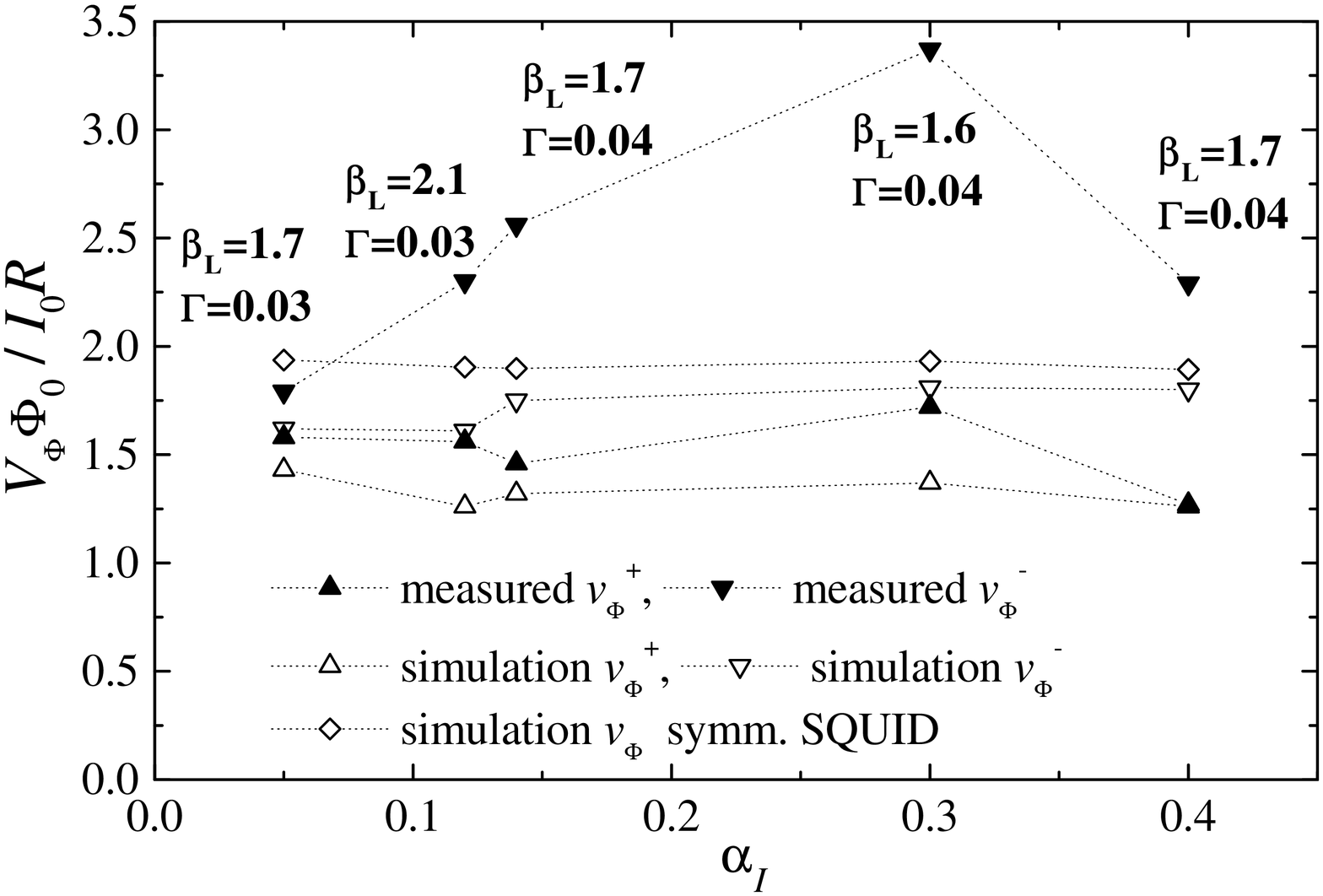}}
\caption{Comparison of
measured and calculated transfer function. The
asymmetry parameter $\alpha_I$ was determined from the
shift of the $V(\Phi)$-characteristics. For comparison
the calculated values of $V_\Phi$ for symmetric dc
SQUIDs are also shown.}
\label{f-Vphi_alpha}
\end{figure}

To test the numerical simulation results we determined
$V_\Phi$ for various YBCO dc SQUIDs (\#2-6 from Table I)
with $\beta_L\approx 2$ and $\Gamma\approx 0.04$ at
$T$=77K. Figure \ref{f-Vphi_alpha} shows that the
simulation (open symbols) predicts a reduction of both,
$v_\phi^+$ and $v_\phi^-$ as compared to the symmetric
case.
For $v_\phi^+$ the experimental data (solid symbols)
lie between the values predicted for the symmetric and
asymmetric case. For $\alpha_I=0.4$ the measured and
calculated values for $v_\phi^+$ are identical and lie
almost a factor of two below the value predicted for
the symmetric SQUID. This supports the idea that at
least part of the discrepancy in $V_\Phi$ between
calculated and measured values can be attributed to
asymmetry. However, most measured values for $v_\phi^-$
are clearly larger than predicted for either the
symmetric or the asymmetric SQUID. The reason for this
deviation is most likely due to the large uncertainty
in the value for $\alpha_R$, which induces an
increasing distortion with increasing $\alpha_R$
\cite{kleiner96,mueller00}.

\begin{figure}[t2bh]
\center{\includegraphics [width=\columnwidth,clip]
{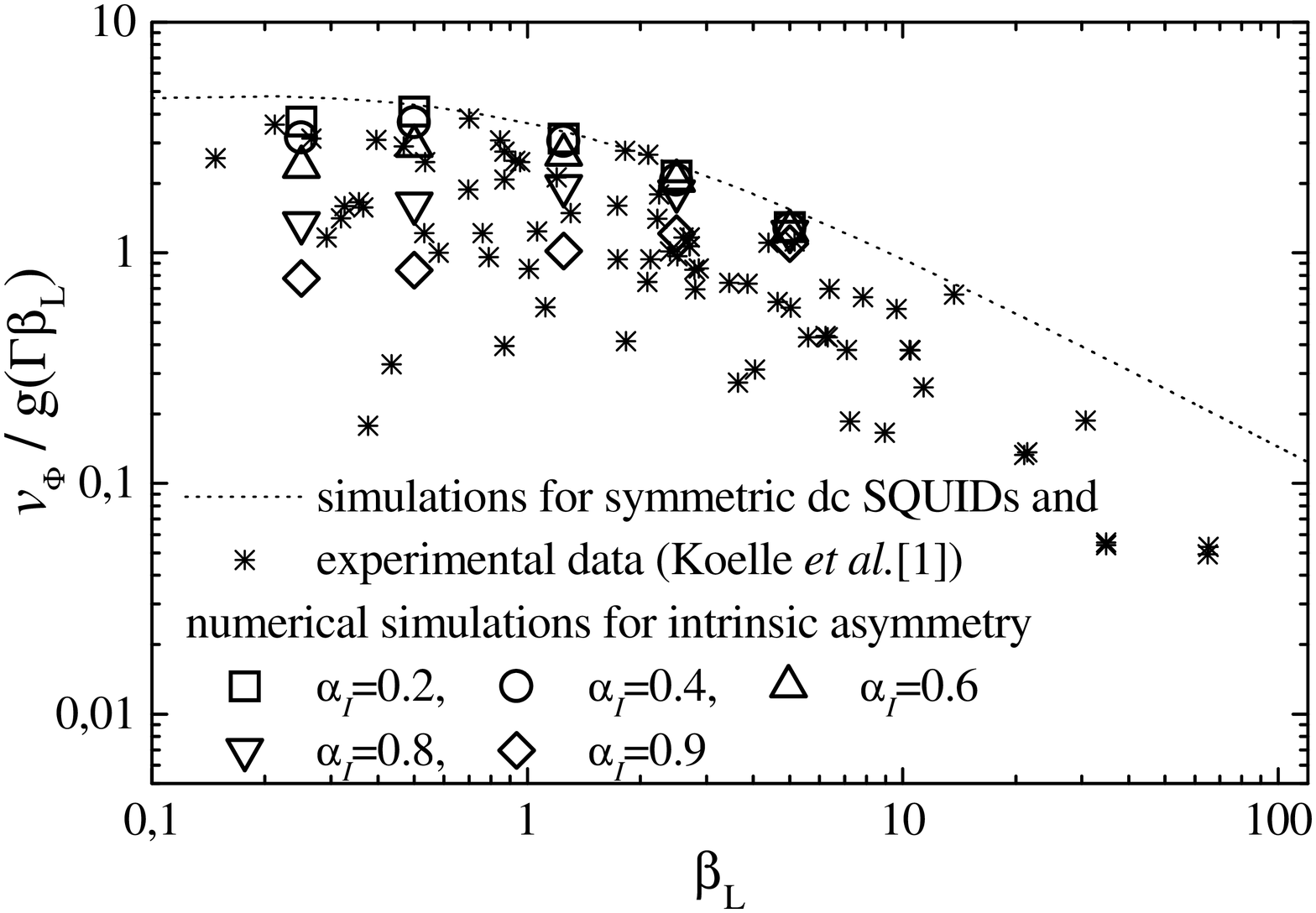}}
\caption{Comparison of calculated and measured
normalized transfer function plotted vs. $\beta_L$.}
\label{f-Vphi_beta-comp}
\end{figure}

We finally note that a clear-cut comparison between
simulation and experimental data for $V_\Phi$ requires
the experimental determination of at least $\alpha_I$,
and preferably also of $\alpha_R$ for a wide variety of
HTS dc SQUIDs. Unfortunately, an extensive collection of
such data does not exist
yet. In order to obtain at least some information on the
importance of asymmetry as a possible source for the
discrepancy in $V_\Phi$ between theory und experiment we
show in Fig.\ref{f-Vphi_beta-comp} the prediction for
symmetric dc SQUIDs (dotted line) compared with
experimental data taken from the literature
\cite{koelle99}.
In addition, the large open symbols
show our simulation results obtained for various values of
$\alpha_I$, assuming intrinsic asymmetry with
$I_0R\propto j_0^{1/2}$. Obviously, the observed deviations
between experiment and simulation for symmetric SQUIDs
cannot be explained by asymmetry for large values of
$\beta_L\gsim5$. For smaller values of $\beta_L$,
however, most of the experimental data lie within the
range of values covered by simulations which take into
account asymmetry in the SQUID, although a large reduction of
$V_\Phi$ due to asymmetry, say by a factor of five
requires a very large $\alpha_I\approx 0.9$.

\section{Conclusions}
\label{conclusions}

We have analyzed the performance of asymmetric HTS dc
SQUIDs both experimentally and by numerical simulation,
with focus on transfer function. Our simulations show
that strong critical current asymmetry which may arise from a
large spread in critical currents in HTS Josephson junctions
can significantly reduce $V_\Phi$ for small $\beta_L\lsim
2$. This observation is important, since optimum
performance requires the realization of small
$\beta_L\approx1$. We wish to stress that the asymmetry,
which is most likely present in almost all HTS dc
SQUIDs, may be one source for the previously found
discrepancy in $V_\Phi$ between experiments and
simulations, however it is not likely that this
asymmetry is the major source of this discrepancy.

\section*{Acknowledgment}
We gratefully acknowledge valuable support from Knut Barthel
and Alex I. Braginski.


%


\end{document}